\documentclass[12pt]{article}
\usepackage{graphicx}
\begin{document}

\title{Dynamical aspects of isotopic scaling}

\author{Martin Veselsky\\
\\
Institute of Physics, Slovak Academy of Sciences,\\
Dubravska cesta 9, Bratislava, Slovakia\\
e-mail: fyzimarv@savba.sk
}

\date{}

\maketitle

\begin{abstract}

Investigation of the effect of the dynamical stage of heavy-ion collisions 
indicates that the  
increasing width of the initial isospin distributions is reflected
by a significant modification of the isoscaling slope for the final 
isotopic distributions after
de-excitation.  For narrow initial distributions, the isoscaling slope
assumes the limiting value of the two individual
initial nuclei while for wide initial isotopic distributions 
the slope for hot fragments approaches the initial value.  
The isoscaling slopes for final cold
fragments increase due to secondary emissions. 
The experimentally observed evolution of the isoscaling parameter 
in multifragmentation of hot quasiprojectiles at E$_{inc}$=50 AMeV, 
fragmentation 
of  $^{86}$Kr projectiles at E$_{inc}$=25 AMeV and multifragmentation of 
target spectators at relativistic energies was reproduced 
by a simulation with the dynamical stage described using the appropriate model 
( deep inelastic transfer and incomplete fusion at the Fermi energy domain 
and spectator-participant model 
at relativistic energies ) and the de-excitation stage described with the 
statistical multifragmentation model. In all cases the isoscaling behavior 
was reproduced by a proper description of the dynamical stage and 
no unambiguous 
signals of the decrease of the symmetry energy coefficient were observed. 

\end{abstract}

\section*{Introduction}

Multifragmentation studies in the recent years highlighted 
the importance of fragment yield ratios which can be 
used to extract thermodynamical observables of the fragmenting system 
such as temperature and chemical potential ( for a review of related methods 
see e.g. \cite{MVIsoTrnd} ). In the context of isotopic 
distributions, the fragment yield ratios represent the details of the 
distribution sensitive to the isospin degrees of freedom. Similar sensitivity 
can be explored globally by investigating the ratio of isotopic yields 
from two processes with different isospin asymmetry, essentially dividing the 
two isotopic distributions in a point-by-point fashion. When employing 
the macro-canonical formula for fragment yields, such a ratio 
will depend on N and Z as follows \cite{TsangIso}

\begin{equation}
     R_{21}(N,Z) = Y_{2}(N,Z)/Y_{1}(N,Z) = C \exp(\alpha N  + \beta Z)
\label{r21isots}
\end {equation}       

with  $\alpha$ = $\Delta \mu_{n}$/T 
and $\beta$ = $\Delta \mu_{p}$/T, where $\Delta \mu_{n}$ and $\Delta \mu_{p}$ 
are the differences in the free neutron and proton chemical potentials, 
respectively,  
of the fragmenting systems. C is an  overall normalization constant.
Alternatively \cite{BotvIso} the N and Z dependence can be expressed as

\begin{equation}
     R_{21}(N,Z) = Y_{2}(N,Z)/Y_{1}(N,Z) 
     = C \exp(\alpha^{\prime} A  + \beta^{\prime} (N-Z) )
\label{r21isobt}
\end {equation}       

thus introducing the parameters which can be related to the isoscalar 
and isovector components of the free nucleon chemical potential since 
$\alpha^{\prime}$ = $\Delta (\mu_{n}+\mu_{p})$/2T 
and $\beta^{\prime}$ = $\Delta (\mu_{n}-\mu_{p})$/2T. 

An exponential scaling of $R_{21}$ with neutron and proton numbers 
was observed experimentally in multifragmentation data 
from collisions of high energy light particles with massive target nuclei 
\cite{BotvIso,Lozhkin} and from collisions between mass symmetric projectiles 
and targets at intermediate energies \cite{TsangIso}. Such exponential 
behavior is called isotopic scaling or isoscaling \cite{TsangIso} 
( the parameters 
$\alpha, \beta, \alpha^{\prime}, \beta^{\prime}$ being referred to as 
isoscaling parameters ). An isoscaling behavior was also reported 
in studies of heavy residues \cite{GSHRIso} and 
in fission data \cite{Fisiso}.  The 
values of the isoscaling parameters were related by several authors 
to various physical quantities such as the symmetry energy 
\cite{TsangIso,BotvIso}, the level of isospin equilibration \cite{GSHRIso} 
and the values of transport coefficients \cite{Fisiso}. 

As demonstrated in the literature, isoscaling appears to be a global feature 
of nuclear reactions and multifragmentation data and the isoscaling 
parameters show sensitivity to both the dynamical and the thermodynamical 
properties of the hot source created in the early stage of the collision. 
It is of interest to clarify to which extent the isoscaling behavior 
is modified by the process of de-excitation in the late stage 
of the reaction and whether the 
dynamical and thermodynamical properties of the hot source 
can be disentangled. 

\begin{figure}[h]                                        

\centering
\includegraphics[width=10.0cm, height=7.0cm ]{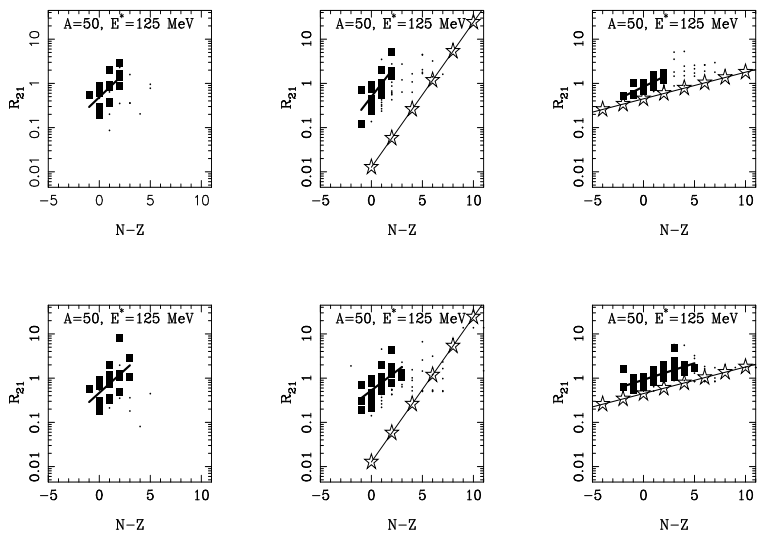}

\caption{
Isoscaling plots after de-excitation by the SMM for nuclei with mass A = 50 
and an excitation energy of 125 MeV. Upper and lower rows represent cold 
and hot partitions, respectively. Squares and thick solid lines - 
isoscaling plots and exponential fits for Z$\leq$6, scattered dots - all Z's. 
The dynamical stage ( stars connected by thin solid lines ) was simulated 
by shifted Gaussians ( N/Z = 1.0, 1.3 ) with three different values of widths  
( $\sigma_Z$ = 0.5, 2.5, 4.5 , see panels from left to right ).  
           }
\label{A50X125}
\end{figure}

\section*{Effect of the de-excitation stage on isoscaling}

To disentangle the dynamical and thermodynamical properties of the hot source 
in the experimental data and, specifically, to determine isoscaling 
properties after the dynamical stage, one can simulate the 
de-excitation process for various initial isotopic distributions
produced in the dynamical stage. Such simulations allow to establish the 
relation between the isoscaling behavior prior to and after the de-excitation 
stage. In the present work the de-excitation stage is simulated 
using the code SMM \cite{SMM}, representing 
the combination of the statistical multifragmentation model ( SMM ) for highly 
excited nuclei with evaporation/fission cascade at lower excitation energies. 
Simulations of the de-excitation stage with the SMM 
proved consistently better than sequential binary decay models, 
especially for the neutron-rich nuclides \cite{MVKrNi} and residues produced 
after the de-excitation of hot nuclei \cite{MVSnAl}. Some discrepancies were 
observed for the yields of a limited set of $\beta$-stable nuclei 
\cite{MVKrNi} close to the projectile, which were overestimated due 
to a low probability for the emission of complex fragments below 
multifragmentation threshold \cite{MVSnAl}. 

The effect of the de-excitation stage was investigated for initial 
( dynamical ) isobaric distributions with three 
masses A = 25, 50, 100, thus covering the typical mass range 
where multifragmentation studies are commonly performed. 
In nuclear reactions, 
the de-excitation 
stage is preceded by the dynamical stage where hot nuclei are formed. 
An open question consists of understanding the isoscaling behavior after 
the dynamical 
stage and how it is modified by the effect of de-excitations. 
A possible way to answer such question is to generate the distributions of hot 
nuclei exhibiting isoscaling and to observe the effect of de-excitation. 
In order to generate the initial dynamical distributions exhibiting isoscaling 
one can explore the well known fact that after dividing two Gaussian 
distributions with equal width and shifted centers, an exponential is obtained 
and thus isoscaling is guaranteed. The logarithmic slope of such exponential 
( commonly referred to as isoscaling parameter ) will thus be determined 
by the centers of the 
two Gaussian distributions, $x_1$ and $x_2$, and their common width, $\sigma$

\begin{equation}
     \alpha = ( x_2 -x_1 )/\sigma^2 .
\label{isogau}
\end {equation}       

Varying of the parameters of these Gaussian distributions allows one  
to vary the initial isoscaling behavior at the early dynamical stage. 
In the present work the 
positions of the centers are fixed and the Gaussian width is used to 
control isoscaling behavior. Such a choice reflects the situation occurring in 
damped nucleus-nucleus collisions where the initial isospin asymmetry 
is not changed dramatically while the width of distribution evolves 
quickly with damping of the initial kinetic energy. 
In the simulations presented here, isoscaling after the dynamical stage 
was simulated using initial isotopic distributions approximated by 
two Gaussians with shifted centers ( N/Z = 1.0, 1.3 ). 
Three different values of common Gaussian widths were used,  
$\sigma_Z$ = 0.5, 1.5, 3.5 for A = 25 and 0.5, 2.5, 4.5 for A = 50, 100 .  
For each mass, the yields of final products were simulated with 
good statistics for hot 
sources with five selected atomic numbers. For other elements, the yields 
of final products were estimated using polynomial interpolations. 
The calculation was carried out for two values of excitation energy,  
E$^{*}$ = 2.5 and 5.0 AMeV, the former being close to multifragmentation 
threshold while the latter corresponding to the region where multifragmentation 
is the main de-excitation mode.

\begin{figure}[h]                                        

\centering
\includegraphics[width=10.0cm, height=7.0cm ]{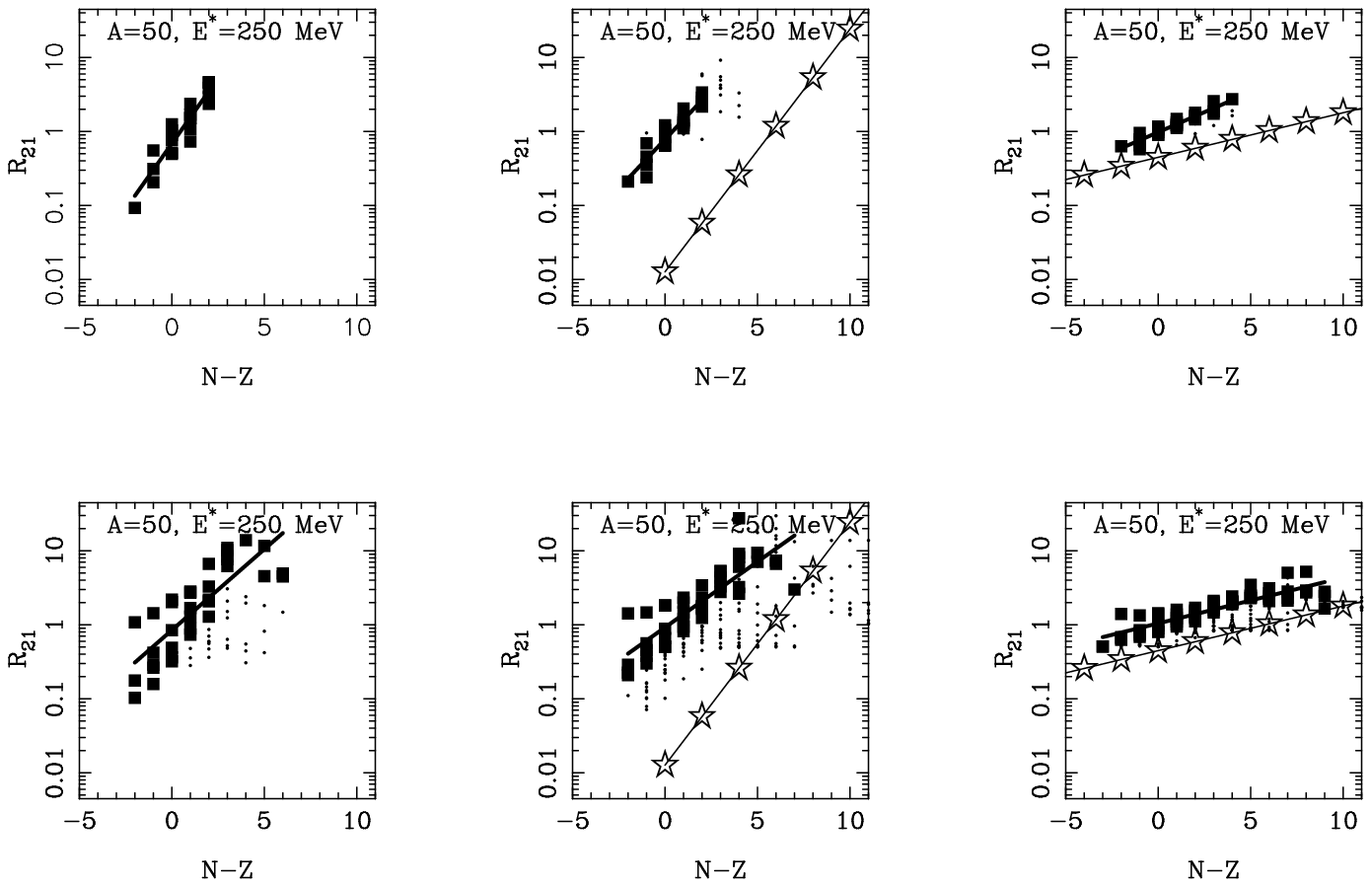}

\caption{
Isoscaling plots after de-excitation by the SMM for nuclei with mass A=50 
and an excitation energy of 250 MeV ( $\sigma_Z$ = 0.5, 2.5, 4.5 ). 
Symbols and lines as in Fig. \ref{A50X125}.  
           }
\label{A50X250}
\end{figure}

In Figs. \ref{A50X125}, \ref{A50X250} are shown results for nuclei 
with mass A = 50 and excitation energies of 125 and 250 MeV, 
respectively. For the narrow 
initial Gaussian distributions ( left columns ) the isoscaling slopes appear 
to be governed by the 
intrinsic effect of de-excitation, while with increasing width 
( middle and right columns ) the 
effect of initial distributions appears to take over. Secondary 
emission leads to a slight increase of the slope due to a lower temperature 
which, according to macro-canonical theory ( Eqs. (\ref{r21isots}) 
and (\ref{r21isobt}) ), enters into the denominator. For the widest 
initial distribution ( right columns ) the isoscaling plot 
in the hot partition appears to 
follow the initial isoscaling plot and the increase of the slope by 
secondary emission determines the final discrepancy. 

\begin{figure}[h]                                        

\centering
\includegraphics[width=10.0cm, height=7.0cm ]{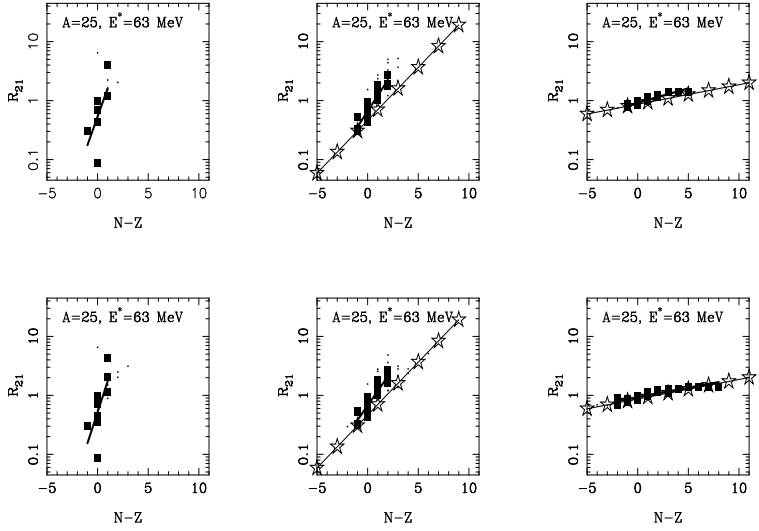}

\caption{
Isoscaling plots after de-excitation by the SMM for nuclei with mass A = 25 
and an excitation energy of 63 MeV ( $\sigma_Z$ = 0.5, 1.5, 3.5 ). 
Symbols and lines as in Fig. \ref{A50X125}.  
           }
\label{A25X63}
\end{figure}

\begin{figure}[h]                                        

\centering
\includegraphics[width=10.0cm, height=7.0cm ]{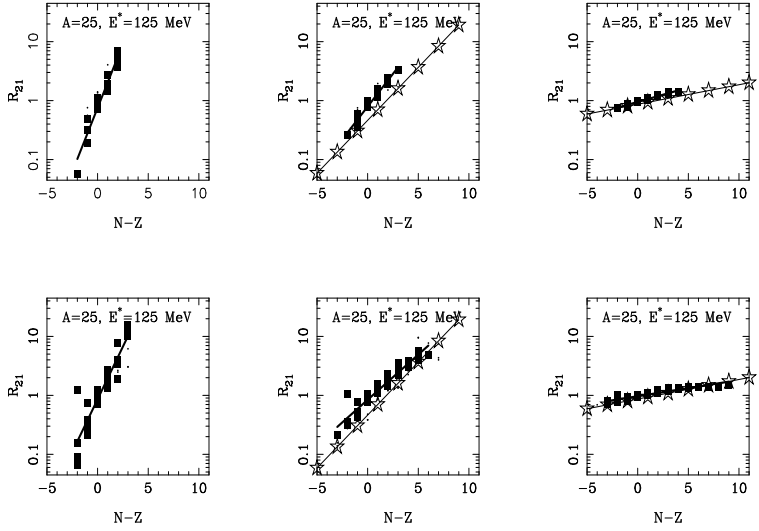}

\caption{
Isoscaling plots after de-excitation by the SMM for nuclei with mass A=25 
and an excitation energy of 125 MeV ( $\sigma_Z$ = 0.5, 1.5, 3.5 ). 
Symbols and lines as in Fig. \ref{A50X125}.  
           }
\label{A25X125}
\end{figure}

Results for nuclei with mass A = 25 and excitation energies 
63 and 125 MeV are shown, respectively, in Figs. \ref{A25X63}, \ref{A25X125}. 
Analogous 
conclusions as for A = 50 can be made. 
However, the dominance of the initial isoscaling width an increase 
of the slope 
by secondary emissions is observed in both middle and right panels, 
indicating that such behavior takes over earlier as the initial width 
increases, as compared to the case of A=50. 
 
\begin{figure}[h]                                        

\centering
\includegraphics[width=10.0cm, height=7.0cm ]{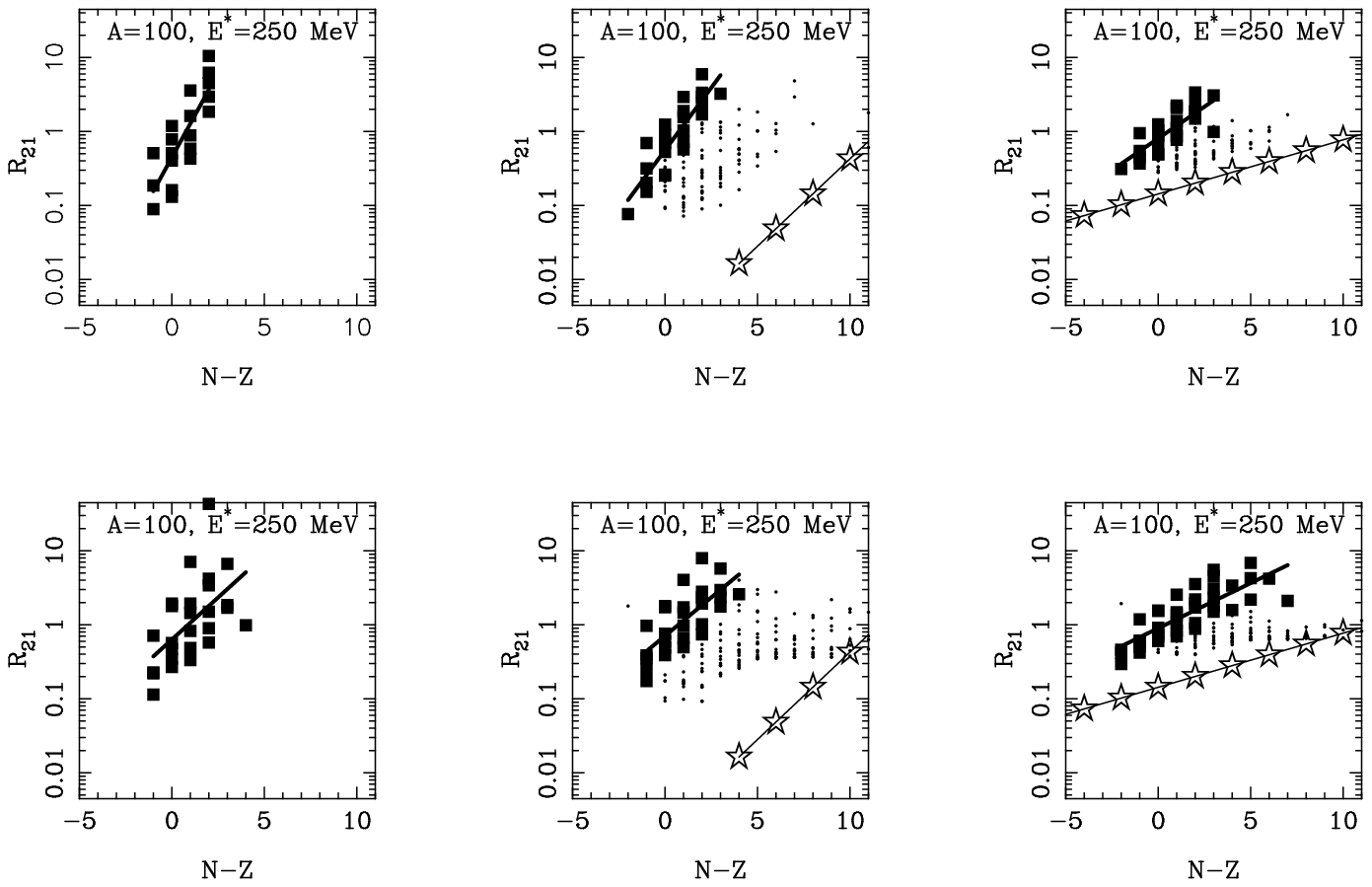}

\caption{
Isoscaling plots after de-excitation by the SMM for nuclei with mass A = 100 
and an excitation energy of 250 MeV ( $\sigma_Z$ = 0.5, 2.5, 4.5 ).  
Symbols and lines as in Fig. \ref{A50X125}.  
           }
\label{A100X250}
\end{figure}

\begin{figure}[h]                                        

\centering
\includegraphics[width=10.0cm, height=7.0cm ]{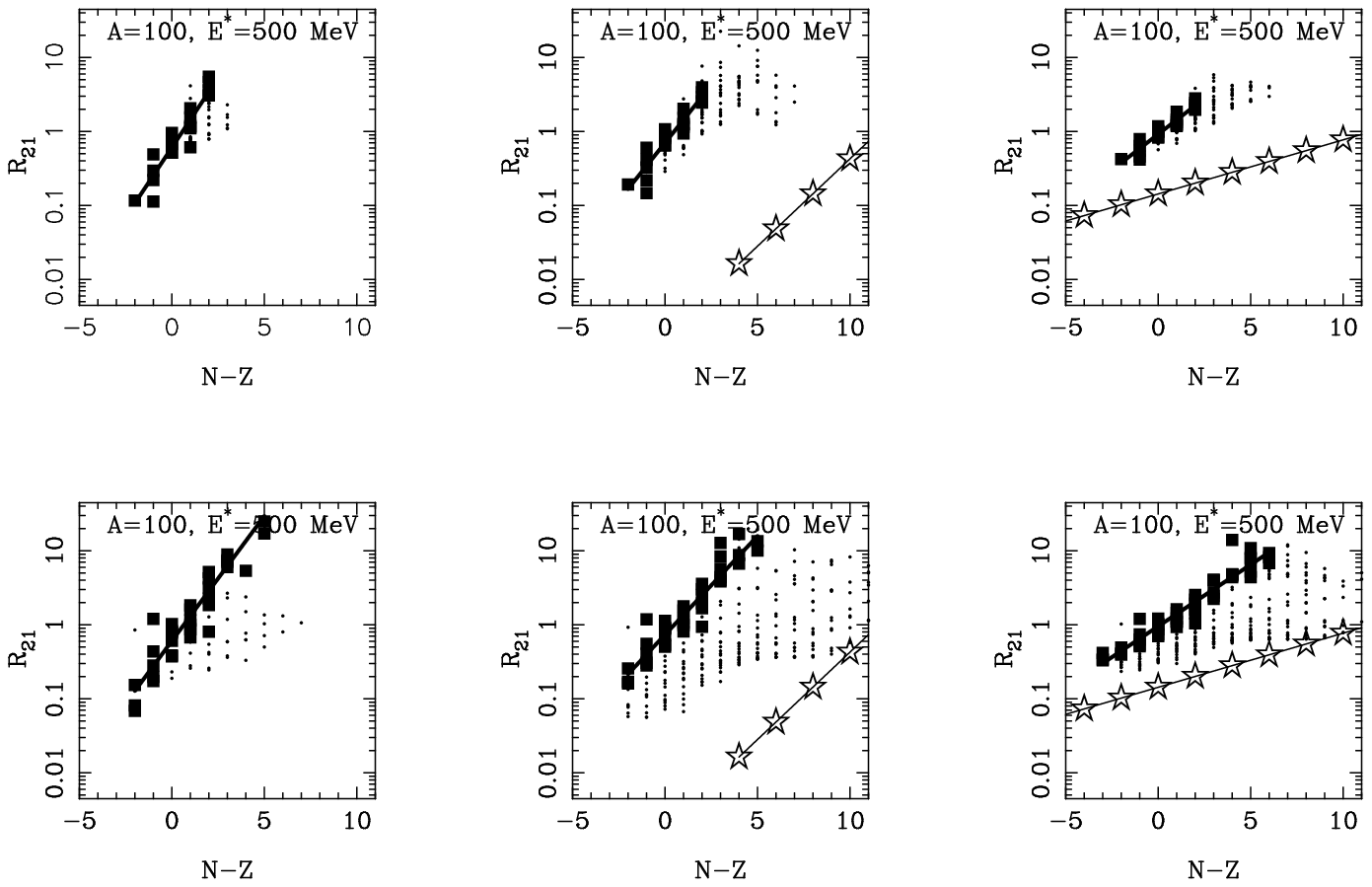}

\caption{
Isoscaling plots after de-excitation by the SMM for nuclei with mass A = 100 
and an excitation energy of 500 MeV ( $\sigma_Z$ = 0.5, 2.5, 4.5 ).  
Symbols and lines as in Fig. \ref{A50X125}.  
           }
\label{A100X500}
\end{figure}

The case of A = 100 ( excitation energies 250 and 500 MeV ) is shown 
in Figs. \ref{A100X250}, \ref{A100X500}. 
The behavior is analogous to the previous cases, but the isoscaling slope 
for hot fragments with Z$\leq$6 in the case of the widest 
initial distribution ( lower right panel ) appears to be larger 
than initial one. This is due to the fact that very isospin-asymmetric nuclei 
are 
produced and the symmetry energy of such light nuclei increases quickly 
and thus increasingly influencing the overall energy balance. However, 
for heavier fragments ( dots ) the isoscaling behavior appears to follow the 
initial distributions better. Such sensitivity to the symmetry 
energy predicted for light fragments originating 
from hot heavy nuclei can in principle 
provide a probe of the symmetry energy coefficient at the hot stage, 
if the 
reaction dynamics leads to wide initial distributions that can be
reconstructed possibly via full 
calorimetry of the hot source or via reliable simulations of the initial stage.    

In general, the increasing width of initial isotopic distributions ( and 
the corresponding decrease of the initial isoscaling slope ) 
is reflected  
by significant modification of the final isoscaling slope after 
de-excitation. For narrow initial Gaussian distributions, the isoscaling slope 
assumes the limiting value fully determined by the details of the 
de-excitation stage. 
For wide initial Gaussian distributions, the isoscaling slope   
for hot fragments approaches the slope of initial isoscaling plots and it 
is thus fully determined by the initial stage. This correspondence is modified 
by secondary emission and the isoscaling slopes for final cold 
fragments are larger possibly due to a corresponding decrease  
of the temperature during secondary emissions. It is noteworthy that 
the width of initial Gaussian distributions induces a decrease of 
the isoscaling parameters comparable to the values, reported in the 
literature \cite{TsangIso,BotvIso}, and explained   
as an effect of a decreasing symmetry energy, 
according to liquid-drop based formula that relates the symmetry 
energy coefficient directly to the isoscaling parameter. However, 
the effect of the dynamical stage and specifically of the width of the initial 
distributions was not considered in the analysis and the estimates 
are based on simulation for individual initial nuclei, which appears 
to be an over-simplified approach.

\section*{Investigation of the dynamical stage in selected reactions}

The investigation presented in previous section suggests that 
the dynamical stage, leading to the evolution of a considerable width 
of the isospin distribution, plays an important role in determining  
the isoscaling behavior of final products. A detailed understanding of 
reaction dynamics is thus necessary to allow disentangling the 
properties of the hot multifragmentation source from the artifacts 
of the reaction dynamics. Three selected cases ( multifragmentation 
of hot quasiprojectiles at incident energy of 50 AMeV, fragmentation 
of a $^{86}$Kr beam at an incident energy of 25 AMeV and multifragmentation of 
target spectators at relativistic energies  ) will be presented 
in this section in order to investigate the effects of reaction 
dynamics on isoscaling in few energy regions for hot 
nuclei with different masses.  

.  
\subsection*{Multifragmentation of projectiles with masses A $\sim$ 25 
at Fermi energies }

Multifragmentation of hot quasiprojectiles with A$\sim$25 
was studied in reactions
$^{28}$Si+$^{124,112}$Sn at projectile energies of 30 and 50 AMeV
\cite{SiSnNExch}. The observed fragment
data \cite{SiSnNExch} provide full information
( with exception of emitted
neutrons ) on the decay of thermally equilibrated hot quasi-projectiles
with known masses ( A = 20 - 30 ), charges, velocities and excitation energies.
A detailed investigation of the reaction mechanism \cite{SiSnNExch}
allowed to establish a dominant reaction scenario.
An excellent description of fragment observables
was obtained using the deep-inelastic transfer ( DIT ) model \cite{DITTGSt}
for the early stage of the collision and the statistical multifragmentation
model ( SMM ) \cite{SMM} for the de-excitation stage. 
The DIT model describes well the
dynamical properties of the reconstructed quasi-projectile such as its center
of mass velocity, excitation energy 
and isospin-asymmetry. Fragment observables
such as multiplicities, charge distributions and mean N/Z values for
a given charge were also reproduced reasonably well \cite{SiSnNExch}.
Thus the reaction mechanism 
can be considered well understood. 
The contribution from non-equilibrium processes
such as pre-equilibrium emission was shown to be weak \cite{SiSnNExch}.
According to successful DIT+SMM simulation, the number of emitted neutrons, 
not detected in the experiment, was between one and two per event
and the underestimation of the excitation energy 
due to undetected neutrons can be estimated to be 
approximately 10--15 \% in the whole range of excitation energies.
The excitation energy dependence of the isoscaling slope, corrected
for the 1/T temperature dependence, exhibits a turning-point at E$^{*}$=4 AMeV
\cite{SiSnIso} which can be interpreted as a signal of the onset of 
separation into an isospin asymmetric dilute phase and an isospin symmetric 
dense phase.  
The onset of the chemical separation is correlated to the onset
of the plateau in the caloric curve, thus signaling
that chemical separation is accompanied by a latent heat. 

\begin{figure}[h]                                        

\centering
\includegraphics[width=7.0cm, height=8.0cm ]{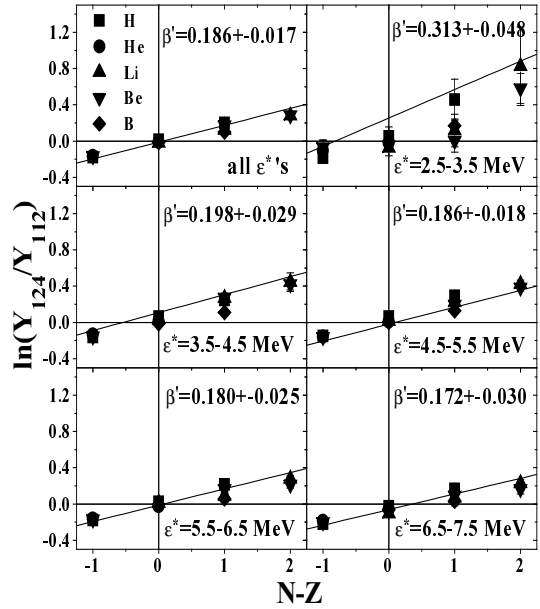}

\caption{
Simulated isoscaling plots ( symbols ) and fits to experimental data 
( lines ) 
from the statistical decay of hot quasi-projectiles in the reactions
$^{28}$Si+$^{124,112}$Sn at incident energies of 50 AMeV. The upper 
left panel corresponds to inclusive 
data while the other panels correspond to the five excitation energy bins. 
           }
\label{Iso50}
\end{figure}

In Fig. \ref{Iso50} are presented simulated isoscaling data ( symbols ) 
from statistical decay of hot quasi-projectiles produced in the reactions
$^{28}$Si+$^{124,112}$Sn at projectile energy 50 AMeV.
The isoscaling plots are presented not only for the inclusive data 
( upper left panel ) but also for 
five bins of excitation energy. The isoscaling slope in the simulations 
depends on the excitation energy
almost identically as in the experimental data, represented by the solid lines. 
The DIT+SMM simulation fully reproduces experimental isoscaling behavior. 
The isoscaling parameters of the initial isospin 
distributions exhibit a similar trend as the final values, in agreement 
with the results of simulation presented in Figs. \ref{A25X63}, \ref{A25X125}.  
The shift between simulated initial isotopic distributions is 
constant in all the excitation energy bins and the evolution is essentially 
determined by their widths that are smaller in the lowest excitation energy 
bins and then depend on excitation energy only weakly. 
Despite the overall success, the simulation does not allow one to extract 
unambiguously the values of the temperature corresponding to isospin 
trends of hot fragment distributions, due to the fact that 
the production of multiple fragments occurs mostly in secondary  
emissions represented by Fermi decay. The formalism of Fermi decay 
is analogous to the multifragmentation model with cold fragment partitions, 
which thus essentially duplicates the multifragmentation 
model with hot fragments used in the SMM. Thus the model does not 
provide an unambiguous equivalent to experimental double-isotope 
ratio or slope temperatures, used in \cite{SiSnIso,SiSnNPA}. The duplicity 
of fragmentation stages in the calculation is consistent with the 
analogous success of the model of sequential binary decay for light 
nuclei, where the proper exploration of available phase space 
appears the most important requirement to successful models.

\subsection*{Fragmentation of $^{86}$Kr beam with $^{124,112}$Sn 
targets at 25 AMeV}

The isoscaling phenomena are not restricted to relatively light fragments
but can be observed also in
heavy residue data, collected at forward angles. Yield ratios
$ R_{21}(A,Z)$ of  projectile residues
from  the reactions $^{86}$Kr+$^{124,112}$Sn at 25 AMeV \cite{KrSn_1}
were investigated and isoscaling behavior was observed 
for each isotopic and isotonic chain. 
The isoscaling 
slopes are constant for residue mass range A = 25 -- 60, corresponding 
to primary events with the maximum observed  excitation energy of 2.2 AMeV. 
The slopes exhibited gradual
decrease with increasing mass of the residues. 
Assuming that the fragmentation occurs at  normal density,
using  C$_{sym}$ = 25 MeV \cite{BotvIso}, the values of isoscaling parameters 
can be used to determine the values of  $ \Delta( N/Z )_{qp} $ 
( where qp means quasi-projectile ) as a function
of the observed residue mass A and charge Z, thus  
demonstrating the evolution of the
N/Z equilibration process in isospin-asymmetric collisions.
The monotonic increase of $ \Delta( N/Z )_{qp} $ with excitation energy
can be understood as a result of the mechanism of nucleon exchange.

\begin{figure}[h]                                        

\centering
\includegraphics[width=14.0cm, height=5.0cm ]{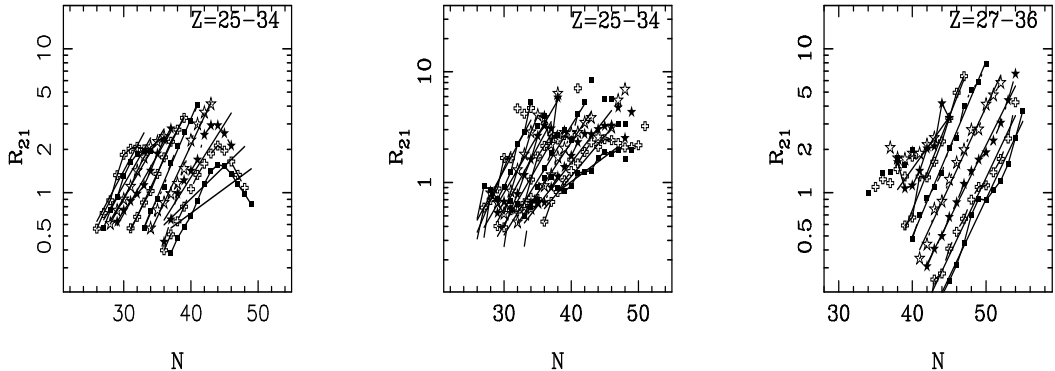}

\caption{
Isoscaling plots for the reactions of $^{86}$Kr+$^{124,112}$Sn at an 
incident energy of 25 AMeV. 
Left panel - simulated data for final fragments, middle panel - experimental 
data \cite{GSKrSn}, right panel - simulated data after dynamical stage. 
The lines represent exponential fits. 
           }
\label{IsoKrSn}
\end{figure}

The left panel of Fig. \ref{IsoKrSn} shows isoscaling plots corresponding to  
the simulations of the reactions of $^{86}$Kr (25AMeV) with $^{124,112}$Sn. 
The used simulation is the same as in \cite{MVKrSn} where it allowed to 
reproduce experimental cross sections for neutron-rich nuclides 
and residues from de-excitation of hot nuclei. Some discrepancies were 
observed in the yields of a limited set of $\beta$-stable nuclei 
close to the projectile, which were overestimated due to a low probability 
for the emission of complex fragments below multifragmentation threshold.  
As a comparison, in the middle panel experimental isoscaling plots are shown. 
For nuclei with Z=25-30 the simulation and experiment lead to a similar 
behavior with constant slopes and consistent values of the 
isoscaling parameters. 
For lower Z's ( Z$\leq$24, not shown in the Fig. \ref{IsoKrSn} ) 
the experimentally observed slopes become even larger than the calculated 
ones, thus eventually implying a physical phenomenon not 
included in the simulations, however this discrepancy  
can be caused by experimental limitations in measuring the yields 
of these elements in the tails of the isotopic distributions 
( as it is discussed in \cite{MVKrSn} ) 
leading to lower widths of the isotopic distributions and thus larger apparent 
values of the isoscaling parameters. 
For heavier nuclei with N$>$44, the simulation leads to a reverse  
trend of the yield ratios toward unity, possibly signaling the onset of a  
reaction mechanism independent of the N/Z of the target, possibly quasi-elastic 
( direct ) few-nucleon transfer taking place in very peripheral collisions. 
The experimental isoscaling behavior for these nuclei shows signs of 
a similar reverted trend, the transition is not as regular as in the 
simulation and the inclusion of the points from this region into the 
exponential 
fits ( lines ) leads to a decrease of the apparent isoscaling slopes.  
Such decrease of the slope of exponential ( "isoscaling" ) fits 
is shown by the lines in the left panel of Fig. \ref{IsoKrSn}, 
despite the very poor quality 
of such fits. Thus the evolution of the apparent exponential slopes  
in both experimental and simulated data suggest a mixing of two components: 
one component very sensitive to the N/Z of the target, possibly 
due to an intense nucleon exchange;  
a second component, insensitive to  the N/Z of the target, possibly 
quasi-elastic few-nucleon exchange. 
This situation is demonstrated in the right panel of  Fig. \ref{IsoKrSn} 
where simulated 
isoscaling plots are shown for the dynamical stage prior to de-excitation. 
The isotopes with Z = 30 - 36 exhibit regular isoscaling behavior, except 
for a structure around N = 50 corresponding to elements close to the 
projectile charge, which can 
be identified with quasi-elastic processes. Despite minor effect 
on isoscaling plots, these points represent a significant portion 
of the reaction cross section and the corresponding wide excitation energy 
distribution leads to a mixing with the data for lighter elements and thus 
to a modification of their isoscaling behavior after de-excitation. 
The discrepancy of the final simulated and experimental isoscaling behavior 
can be possibly attributed to an underestimated probability 
for the emission of complex fragments below multifragmentation threshold 
in the SMM. 

\begin{figure}[h]                                        

\centering
\includegraphics[width=7.0cm, height=7.0cm ]{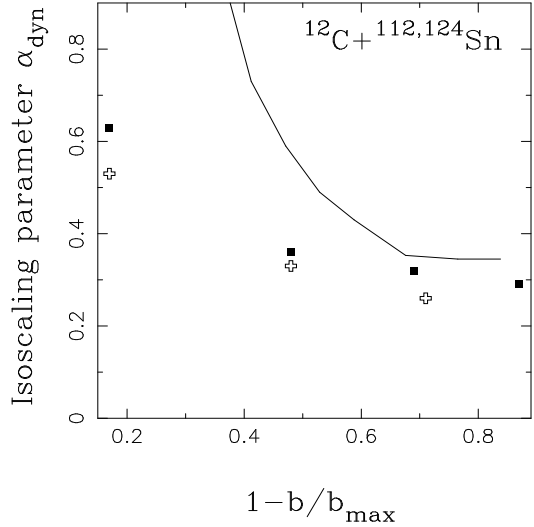}

\caption{
Evolution of the isoscaling parameter at the dynamical stage 
as a function of the centrality 
for the reactions $^{12}$C+$^{112,124}$Sn at relativistic energies. 
Solid line - results of simulations. 
Symbols - experimental data \cite{LeFevre}. 
           }
\label{IsoSnC}
\end{figure}

\subsection*{Multifragmentation of target spectators 
at relativistic energies}

The dominant reaction mechanism at the relativistic energies is represented 
by the  
spectator-participant model where a hot region is formed in 
the participant zone ( zone of geometric overlap ) 
while the spectator regions are colder. 
These spectators can be warm enough to undergo 
multifragmentation. The isoscaling behavior in multifragmentation 
of target spectators was investigated in the literature  
\cite{Lozhkin,LeFevre} and a dependence of the isoscaling parameters 
on the centrality of the collision was observed \cite{LeFevre}. The value 
of the isoscaling parameters was related to the symmetry energy 
\cite{TsangIso,BotvIso} and the decrease of the  
symmetry energy coefficient was reported \cite{LeFevre}. However, based on the 
conclusions of the previous section, the effect of the reaction dynamics, 
specifically of the evolving width of the mass and charge distributions, can 
be considered as an alternative interpretation. 
The volume and thus the most probable mass $A_{TS}^{abr}$ 
of the target spectator can be estimated as a function of the impact parameter 
using the model of geometrical abrasion \cite{Gosset}. 
The number of nucleons in the 
spectator zone $A_{TS}$ can be estimated 
according to the binomial distribution  

\begin{equation}
     P(A_{TS}) = (^{A_{T}}_{A_{TS}}) (\frac{A_{TS}^{abr}}{A_{T}})^{A_{TS}}
     (1-\frac{A_{TS}^{abr}}{A_{T}})^{A_{T}-A_{TS}}
\label{a1binom }
\end {equation}       

where $A_{T}$ is the target mass number.   
The charge of the spectator $Z_{TS}$ can be determined as \cite{Fried}

\begin{equation}
\label{ffried}
P(Z_{TS}) = \frac{(^{Z_{T}}_{Z_{TS}})(^{N_{T}}_{N_{TS}})}{(^{A_{T}}_{A_{TS}})}
\end{equation}

where $Z_{T}$ is the target mass number and $N_{TS}$ and $N_{T}$ are the 
neutron numbers of the target spectator and the target, respectively. 
The excitation energy of the target spectator can be estimated, 
according to \cite{Gaim}, 
as proportional to the number of abraded nucleons 
with the proportionality factor 27 MeV, which was found to be 
consistent with experimental data \cite{27MeV}. 

Fig. \ref{IsoSnC} shows the evolution of the isoscaling 
parameter after the dynamical stage ( solid line ), obtained using 
centers and widths of simulated fragment   
distributions from the abrasion calculation, as a function of centrality 
for the reactions $^{12}$C+$^{112,124}$Sn at relativistic energies. 
Symbols show the experimental points reported in \cite{LeFevre} 
for incident energies of 300 and 600 AMeV. The calculation is capable 
to reproduce the 
experimental trend without any assumptions about the decrease of the symmetry 
energy coefficient. The discrepancy at low centrality ( small 
excitation energy ) can be explained by the  
asymptotics of the experimental points ( after the de-excitation stage ) 
for the width approaching zero where the isoscaling parameter 
assumes a value determined by the intrinsic properties of the de-excitation. 
In central collisions the isoscaling parameter exhibits analogous 
saturation at a somewhat higher value ( by 10 - 15 \% ) than the ones reported 
for the experimental data.  The calculated widths of the target spectator 
isospin distributions at saturation are similar to the situation in the middle 
panel of Figs. \ref{A100X250}, \ref{A100X500}, where 
the initial ( dynamical ) isoscaling parameter reflects the initial value 
with remaining discrepancies due to secondary emissions. In the present case 
the excitation energy exceeds 11 AMeV and the effect of secondary emissions 
on the slope parameter in the test calculations was within statistical 
errors ( not exceeding 10 \% ). 
Thus after taking secondary emissions into account the 
discrepancy will raise to the level of about 20 \%, resulting in the
corresponding decrease of the apparent 
symmetry energy coefficient from 25 MeV to about 20 MeV. 
Such a decrease is comparable with the 
uncertainty in the chosen initial values 
of the symmetry energy coefficient in the models of nuclear ground state 
properties ( with the value of 23 MeV being commonly used ). 
The decrease of the apparent symmetry energy coefficient can be further caused 
by other dynamical phenomena not included 
in the model such as emission at the pre-equilibrium stage 
( leading to a decrease of the shift between the centroids and to an 
additional increase of the widths of the isotopic distributions ).  
The apparent value of the symmetry energy coefficient around 20 MeV can thus 
hardly be interpreted as a signal of a significant decrease of the nuclear 
symmetry energy. In any case it is much less significant than it was reported 
in \cite{LeFevre} ( up to factor of 6 after taking into account secondary 
emissions ), where the effect of the dynamical
evolution of the initial isotopic distributions ( the width in particular ) 
was 
not considered. The model description of the dynamical evolution presented 
here thus allows one to reproduce the reported discrepancy, 
with the 
remaining discrepancies on the apparent symmetry energy coefficient being not 
significant enough to be declared as an unambiguous signal. 

\section*{Summary and conclusions}

Investigation of the effect of the dynamical stage of heavy-ion collisions 
established that the 
increasing width of the initial isotopic distributions induces 
a significant modification of the isoscaling slopes after
the de-excitation stage.  For narrow isotopic distributions, the isoscaling 
slope assumes the limiting value for two individual
initial nuclei which is fully determined by the details of the de-excitation. 
For wide initial isotopic distributions, the isoscaling slope 
for hot fragments approaches the initial isoscaling slope and
it is thus fully determined by the initial stage. 
This correspondence is modified
by secondary emissions and the isoscaling slopes for final cold
fragments are larger by an amount possibly corresponding to a lowering
temperature during secondary emissions. The decrease of
the isoscaling parameters, caused by the increase of the width 
of initial Gaussian distributions, is comparable in magnitude to the values, 
reported in the literature as an effect of the decrease of the symmetry 
energy. 
The experimentally observed evolution of the isoscaling parameter in the 
statistical decay of hot quasiprojectiles from the reactions
$^{28}$Si+$^{124,112}$Sn at projectile energy 50 AMeV is reproduced 
by a simulation with the dynamical stage described by the  
deep inelastic transfer model and the de-excitation stage described using the 
statistical multifragmentation model. 
The evolution of the apparent isoscaling 
slopes in both experimental and simulated data 
for projectile residues 
from  the reactions of $^{86}$Kr+$^{124,112}$Sn at incident energies 
of 25 AMeV
suggests a mixing of two components,
one sensitive to the N/Z of the target, possibly due to 
an intense nucleon exchange,
and a second component due to a quasi-elastic few-nucleon exchange, 
almost insensitive to the N/Z of the target. 
The discrepancy between the final simulated and experimental isoscaling 
behavior 
can be possibly attributed to the absence of a mechanism 
for the emission of complex fragments below multifragmentation threshold 
in the SMM. 
The decrease of the isoscaling parameter in the multifragmentation of 
target spectators in central collisions was reproduced using the simulation 
using the spectator-participant model for the dynamical stage and the SMM 
model for the de-excitation 
stage. In all cases the isoscaling behavior was reproduced by a proper 
description of the dynamical stage and no unambiguous signals on the decrease 
of the symmetry energy coefficient were observed. 

The author acknowledges L. Tassan-Got for providing his DIT code and 
A.S. Botvina for providing his SMM code. 
This work was supported through grant of Slovak Scientific Grant Agency
VEGA-2/5098/25.

\end{document}